\documentclass[aps,pra,twocolumn,amsmath,showpacs,superscriptaddress]{revtex4}

\usepackage{graphicx}
\usepackage{amssymb}
\usepackage{mathrsfs}

\newcommand{\lit}{\mbox{Li$_{2}$\ }}
\newcommand{\litp}{\mbox{Li$_{2}^{+}$\ }}
\def\mathbi#1{\textbf{\em #1}}

\begin{document}

\title{Phase information revealed by interferences in the ionization of rotational wave packets}

\author{Eric Charron}

\email{Eric.Charron@ppm.u-psud.fr}

\affiliation{
Laboratoire de Photophysique Mol\'{e}culaire du CNRS,\\
Universit\'{e} Paris-Sud 11, B\^{a}timent 210,
91405 Orsay Cedex, France.}

\author{Maurice Raoult}

\email{Maurice.Raoult@lac.u-psud.fr}

\affiliation{
Laboratoire Aim\'{e} Cotton du CNRS,\\
Universit\'{e} Paris-Sud 11, B\^{a}timent 505,
91405 Orsay Cedex, France.}

\date{\today}

\begin{abstract}
Time-resolved photoelectron spectra are proposed for the measurement of classical information recorded in the quantum phases of a molecular
rotational wave packet. Taking \mbox{Li$_{2}$} as a prototypical system, we show that an interference arises from the electron-nuclei entanglement
induced by the molecular anisotropy. This phenomenon is used to transfer the information that was stored initially in the nuclear rotational
degree of freedom into the electronic degree of freedom.
\end{abstract}


\maketitle

\section{Introduction}
\label{sec:Introduction}

In parallel with quantum measurement and quantum entanglement, quantum interference is one of the usual ingredients of quantum communication and computing~\cite{QI}. The overall speedup expected from quantum computers originates in part from the fact that the input state of the computation can be chosen as a coherent superposition of all possible classical inputs. A sequence of unitary operations is applied to this initial state in order to perform the computation. This approach processes the data in a massively parallel way, and the result of the computation usually depends on the interference between the various paths followed by all possible initial states. The real power of quantum interference is revealed when bringing together all different components of the wave function in a single step~\cite{Lloyd99}. Even if not sufficient for reliable and efficient quantum computation~\cite{Lloyd99,Divincenzo00}, the buildup of a quantum interferometer is of course necessary for the construction of a quantum computer. In this respect various physical implementations have been proposed to test the principles of quantum computing, including nuclear magnetic resonance~\cite{Jones01}, trapped ions~\cite{TI}, cavity quantum electrodynamics~\cite{QED}, Josephson junctions~\cite{JJ} and neutral atoms~\cite{NA}.

Matter-wave interferometry using molecular systems is an active field of research~\cite{MWI} with potential applications for quantum information processing. In this article, we show that in a molecular system quantum interferences can be used in combination with electron-nuclear entanglement to extract classical information initially stored in a rotational wave packet. This approach is inspired by a set of beautiful experiments performed on Rydberg atom data registers~\cite{Bucksbaum1,Bucksbaum2}, and on \lit in the context of femtochemistry~\cite{Leone1} and for coherent control~\cite{Leone2,Leone3}. This last system has also been used for the implementation of various simple quantum algorithms~\cite{Leone4}.

\section{The molecular system and the laser fields}
\label{sec:The molecular system and the laser fields}

\begin{figure}[!ht]
\includegraphics[width=8cm,clip]{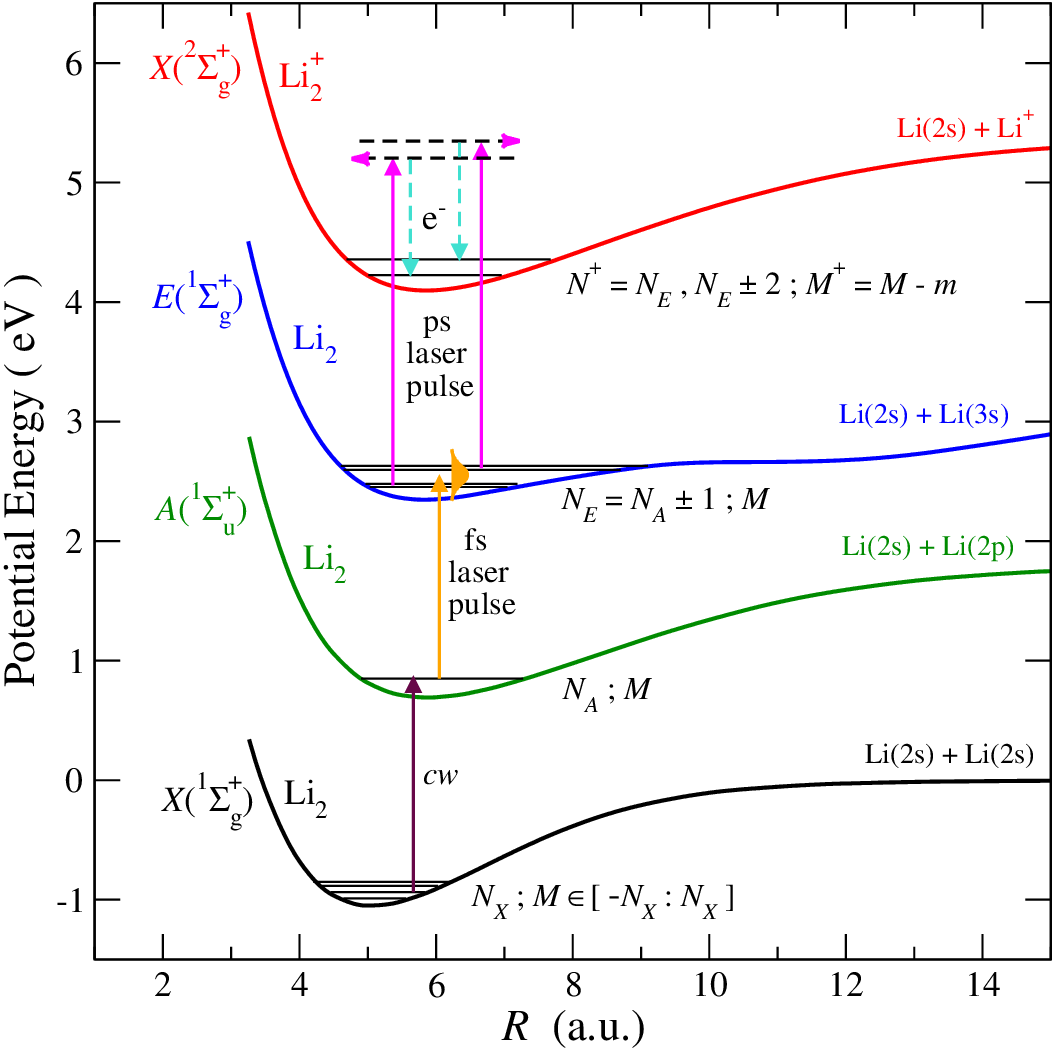}
\caption{
\label{fig:pot}
(Color online)
Three step ``preparation-pump-probe'' scheme proposed to store classical information in the phases of a rotational wave packet in the $E\rm\left(^{1}\Sigma_{g}^{+}\right)$ electronic state of the \lit molecule, and to read it in the photoelectron spectrum recorded in the last step which consists of a picosecond photoionization experiment. The potential energy curves of the $X$, $A$ and $E$ electronic states of \lit are shown as a function of the internuclear distance $R$ as black, green and blue solid lines respectively. The potential curve associated with the ground electronic state of \litp is shown as a red solid line. Their associated molecular rotational quantum numbers are denoted by $N_X$, $N_A$, $N_E$ and $N^+$, while  $M$ and $M^+$ denote their projection on the electric field polarization axis.}
\end{figure}

The present theoretical study is based on the usual three-step photoionization scheme used in most recent experiments~\cite{Leone1,Leone2}, as shown in Figure~\ref{fig:pot}.

A {\it cw\/} linearly polarized laser pulse first excites a {\it single\/} transition from the ground electronic state $X\rm\left(^{1}\Sigma_{g}^{+}\right)$ of \lit to a pure rovibrational $\left(v_A,N_A\right)$ level on the first excited $A\rm\left(^{1}\Sigma_{u}^{+}\right)$ electronic potential curve. From this unique launch state an ultrafast laser pulse of linear polarization excites a superposition of rovibrational levels $\left(v_E,N_{E}=N_A\pm1\right)$ on the $E\rm\left(^{1}\Sigma_{g}^{+}\right)$ electronic potential. We assume here that this pump pulse has been sent through a dispersion-free pulse shaper. This kind of technique has already been used with \lit\cite{Leone2}. A phase mask is then used to control accurately the phases of the various components of the nuclear wave packet created on the $E$ electronic state potential.

Each possible value of the projection \mbox{$M \in [-N_{X} : N_{X}]$} of the initial state rotational quantum number on the electric field polarization axis is equiprobable. We therefore perform a separate calculation for each initial value of $M$, and the total photoelectron signal is calculated by averaging our results over this initial state quantum number (section~\ref{sec:photoelectron spectra}). Note that the three excitation steps shown in Figure~\ref{fig:pot} are performed with the same electric field linear polarization, and the value of $M$ is thus identical in all electronic states of \lit\!\!.

In analogy with the study performed on Rydberg atom data registers~\cite{Bucksbaum1,Bucksbaum2}, we associate the phases imprinted in the coherent superposition of rovibrational levels $\left(v_E,N_{E}\right)$ with information of classical nature. For example, if a single vibrational level $v_E$ is populated, the phase shift $\Delta\varphi$ between the two rotational components $N_{E}=N_A+1$ and $N_{E}=N_A-1$ is fixed at the value 0 or $\pi$. If $\Delta\varphi=0$ the two components are in phase, and this corresponds arbitrarily to the storage of the integer $n=1$. If $\Delta\varphi=\pi$ the two components are out of phase, and this is interpreted as the storage of the integer $n=0$~\cite{Note1}.

With two vibrational levels $v_E$, one can store in a similar way the combinations 00, 01, 10 and 11, corresponding to the integers $n=0$, 1, 2, and 3 where the first binary digit is determined by the phase shift between the rotational components of the first vibrational level while the second digit is controlled similarly by the second vibrational level. This can be generalized, and with $n_v$ vibrational levels one can store all integers up to $n=2^{n_v}-1$. Encoding this type of classical information in a quantum system is of course a very challenging task, and efficient schemes are necessary to reveal this information in the last experimental step. We propose here to read this information with the help of a picosecond ionizing pulse, as shown in Figure~\ref{fig:pot}.

This last laser pulse, the probe pulse, has a bandwidth $\Delta\omega$ which encompasses the two rotational components associated with the various vibrational levels without mixing the different vibrational states during the ionization process, {\em i.e.\/}
\begin{equation}
2B_{\rm rot}\left(2{N_A}+1\right)\; \ll \;\hbar\,\Delta\omega\; \ll \;\hbar\,\omega_{\rm vib}~,
\end{equation}
where $B_{\rm rot}$ and $\omega_{\rm vib}$ are the molecular rotational constant and the vibrational frequency in the $E$-state. Section~\ref{sec:Results and discussion} will show that the encoded number can be revealed by an interference effect modifying the time-resolved photoelectron spectrum $P(\varepsilon)$ which exhibits a series of peaks characteristic of the recorded integer.

\section{Theoretical approach}
\label{sec:Theoretical approach}

\subsection{The molecular basis set}
\label{sec:molecular basis}

We follow the dynamics of the \lit molecule by propagating in time its associated electro-nuclear wave function \mbox{$\Psi(\vec{\mathbi{r}}_{\mathbi{\!e\,}},\vec{\mathbi{R}},t)$} decomposed in two parts 
\begin{equation}
\label{eq:wf}
\Psi(\vec{\mathbi{r}}_{\mathbi{\!e\,}},\vec{\mathbi{R}},t)
=
\Psi_{E}(\vec{\mathbi{r}}_{\mathbi{\!e\,}},\vec{\mathbi{R}},t)
+
\Psi_{+}(\vec{\mathbi{r}}_{\mathbi{\!e\,}},\vec{\mathbi{R}},t)
\end{equation}
corresponding to the $E$-electronic state of \lit and to \mbox{(\litp + e$^{-}$)} respectively. The coordinates of all electrons are denoted by the vector \mbox{$\vec{\mathbi{r}}_{\mathbi{\!e\,}}$}, and the vector \mbox{$\vec{\mathbi{R}} \equiv (R,\,\hat{\mathbi{\!R}}\,)$} represents the internuclear coordinate. We now separate the global electronic coordinate $\vec{\mathbi{r}}_{\mathbi{\!e\,}}$ of all electrons into the coordinate $\vec{\mathbi{r}}_{\mathbi{\!c\,}}$ of the core electrons and the coordinate $\vec{\mathbi{r}}$ of the ionized electron. We then write the electro-nuclear wave functions $\Psi_{E}$ and $\Psi_{+}$ as the following Born-Oppenheimer expressions
\begin{subequations}
\label{eq:wfBO}
\begin{eqnarray}
\label{eq:wfEBO}
\Psi_{E}(\vec{\mathbi{r}}_{\mathbi{\!e\,}},\vec{\mathbi{R}},t)
& = &
\psi_{E}(\vec{\mathbi{R}},t)\,
\Phi_{E}(\vec{\mathbi{r}}_{\mathbi{\!e\,}}|R)\\
\label{eq:wf+BO}
\Psi_{+}(\vec{\mathbi{r}}_{\mathbi{\!e\,}},\vec{\mathbi{R}},t)
& = &
\psi_{+}(\vec{\mathbi{r}},\vec{\mathbi{R}},t)\,
\Phi_{+}(\vec{\mathbi{r}}_{\mathbi{\!c\,}}|R)
\end{eqnarray}
\end{subequations}
where \mbox{$\Phi_{E}(\vec{\mathbi{r}}_{\mathbi{\!e\,}}|R)$} and \mbox{$\Phi_{+}(\vec{\mathbi{r}}_{\mathbi{\!c\,}}|R)$} denote the electronic wave functions associated with the $E$-state of \lit and with the ground electronic state of \litp respectively.

The $E$-electronic state ($^1\Sigma_g^+$ symmetry) is now considered as a 3s$\sigma$ Rydberg state, and the electronic wave function \mbox{$\Phi_{E}(\vec{\mathbi{r}}_{\mathbi{\!e\,}}|R)$} is expressed in the molecular frame (Hund's case (b) representation) as
\begin{equation}
\label{eq:PhiE}
\Phi_{E}(\vec{\mathbi{r}}_{\mathbi{\!e\,}}|R)
=
\phi_{E}(r,\vec{\mathbi{r}}_{\mathbi{\!c\,}}|R)\,Y_{00}(\,\hat{\mathbi{\!r}}\,)
\end{equation}
On the other hand, the electronic part of \mbox{$\psi_{+}(\vec{\mathbi{r}},\vec{\mathbi{R}},t)$} is expressed in the laboratory frame (Hund's case (d) representation) as
\begin{equation}
\label{eq:psi+}
\psi_{+}(\vec{\mathbi{r}},\vec{\mathbi{R}},t)\!=\!\!
\int\!\!d\varepsilon\sum_{\ell,m}\psi^{+}_{\ell m}(\varepsilon,\vec{\mathbi{R}},t)\,
\phi_{\ell}(\varepsilon,r|R)\,Y_{\ell m}(\,\hat{\mathbi{\!r}}\,)
\end{equation}
where \mbox{$\phi_{\ell}(\varepsilon,r|R)$} is the electronic continuum wave function at energy $\varepsilon$ and $(\ell,m)$ denote the electron angular momentum and its projection in the laboratory frame. Note that even though the one-photon transition considered here from the $E$-electronic state will result in the ejection of a $p$-electron with $\ell=1$, we keep the double sum over $\ell$ and $m$ in the following for the sake of generality. Various similar approaches have already been used for the calculation of femtosecond time-resolved photoelectron angular distributions~\cite{FTPD}.

In the present study the photoionization step is performed by a picosecond laser pulse of linear polarization. The following selection rule therefore applies for the projections $M$ and $M^{+}$ of the molecular rotational angular momenta of \lit and \litp on the polarization axis
\begin{equation}
\label{eq:M+}
M=M^{+}+m
\end{equation}
For any given initial projection $M$, an unambiguous relation therefore relates $M^{+}$ and $m$. Hence the quantum number $M^{+}$ is replaced in the following by \mbox{$M-m$}.

In our time-dependent approach a propagation is performed independently for each initial value of $M$, and the angular parts of the nuclear wave packets \mbox{$\psi_{E}(\vec{\mathbi{R}},t)$} and \mbox{$\psi^{+}_{\ell m}(\varepsilon,\vec{\mathbi{R}},t)$} are thus expressed as the following expansions
\begin{subequations}
\label{eq:psiE+}
\begin{eqnarray}
\label{eq:psiE}
\psi_{E}(\vec{\mathbi{R}},t)
& \!=\! & \!\sum_{N_{E}} \psi^{N_{E}}_{M}(R,t) \, {\cal D}^{{N_{E}}^{\,*}}_{M,0}(\,\hat{\mathbi{\!R}}\,)\\
\label{eq:psi+lm}
\psi^{+}_{\ell m}(\varepsilon,\vec{\mathbi{R}},t)
& \!=\! & \!\sum_{N^{+}} \psi^{N^{+}\!,\,\ell}_{M-m}(\varepsilon,R,t) \, {\cal D}^{N^{+\,*}}_{M-m,0}(\,\hat{\mathbi{\!R}}\,)
\end{eqnarray}
\end{subequations}
in terms of the normalized Wigner rotation matrices ${{\cal D}^{N^{\,*}}_{M,\Lambda}}(\,\hat{\mathbi{\!R}}\,)$ which verify
\begin{equation}
\label{eq:Dwigner}
\int\!d\,\hat{\mathbi{\!R}}\; {\cal D}^{N^{\,*}}_{M,\Lambda}(\,\hat{\mathbi{\!R}}\,)\, {\cal D}^{N'}_{M',\Lambda'}(\,\hat{\mathbi{\!R}}\,)
= \delta_{NN'}\,\delta_{MM'}\,\delta_{\Lambda\Lambda'}
\end{equation}
where $\delta$ stands for the Kronecker delta symbol~\cite{Zare}.

\subsection{The laser-molecule interaction}
\label{sec:interaction}

The initial electro-nuclear wave packet~(\ref{eq:wf}) is prepared at time $t=0$ from a ``source level'' $\left(v_X,N_X\right)$ as a coherent superposition of several rovibrational levels \mbox{$\left(v_E,N_{E}\right)$} in the two-step process depicted in Figure~\ref{fig:pot}. The initial components of the expansions~(\ref{eq:psiE+}) are thus given by
\begin{subequations}
\label{eq:wf0}
\begin{eqnarray}
\label{eq:wfE0}
\psi^{N_{E}}_{M}(R,0) & = & c_{N_E,M}\,\sum_{v_E}\;\mathrm{e}^{i\varphi_{v_E,N_E}}\;\chi_{v_E,N_E}(R)\;\;\;\;\;\;\\
\label{eq:wf+0}
\psi^{N^{+}\!,\,\ell}_{M-m}(\varepsilon,R,0) & = & 0
\end{eqnarray}
\end{subequations}
where
\begin{eqnarray}
\label{eq:coef0}
c_{N_E,M}
& = &
\sqrt{2N_{E}+1}\;(2N_{A}+1)\,\sqrt{2N_{X}+1}\nonumber\\
& \times &
\left(
\begin{array}{ccc}
N_{A} & 1 & N_{X}\\
M     & 0 & -M
\end{array}
\right)
\left(
\begin{array}{ccc}
N_{A} & 1 & N_{X}\\
0     & 0 & 0
\end{array}
\right)\nonumber\\
& \times &
\left(
\begin{array}{ccc}
N_{E} & 1 & N_{A}\\
M     & 0 & -M
\end{array}
\right)
\left(
\begin{array}{ccc}
N_{E} & 1 & N_{A}\\
0     & 0 & 0
\end{array}
\right)
\end{eqnarray}
The rovibrational eigenstates $\chi_{v_E,N_E}(R)$ of total energy $E(v_E,N_E)$ are solution of the time-independent Schr\"odinger equation
\begin{equation}
\label{eq:TISE}
\hat{\cal H}^{E}_{N_{E}}\;\chi_{v_E,N_E}(R) = E(v_{E},N_{E})\;\chi_{v_E,N_E}(R)
\end{equation}
where the nuclear Hamiltonian $\hat{\cal H}^{E}_{N_{E}}$ is given by
\begin{equation}
\label{eq:nuchamE}
\hat{\cal H}^{E}_{N_{E}} =
-\frac{\hbar^2}{2\mu}\left[\frac{\partial^2}{\partial R^2}-\frac{N_{E}(N_{E}+1)}{R^2}\right] + V_{E}(R)
\end{equation}
We assume that the $E$-state coherent superposition of rovibrational levels has been created with a femtosecond laser pulse which has been sent thought a phase mask such that the phases $\varphi_{v_E,N_E}$ associated with each level can be controlled. The binary information is then stored in the phase differences \mbox{$\Delta\varphi_{v_E}=\varphi_{v_E,N_A+1}-\varphi_{v_E,N_A-1}$} between the two rotational components \mbox{$N_E = N_A \pm 1$} of each vibrational level $v_E$. In addition we have also assumed that an amplitude mask has been used to compensate the Franck-Condon factors which normally govern the $A \leftarrow X$ transition. The initial populations of the different levels in Eq.~(\ref{eq:wfE0}) have therefore been taken as independent of the vibrational quantum number $v_{E}$.

The molecule is then submitted to an ionizing laser pulse associated with the linearly polarized electric field
\begin{equation}
\label{eq:Electric field}
\vec{\mathbi{E}}(t)= E_0\,f(t)\cos(\omega t)\;\hat{\mathbi{\!e}}
\end{equation}
where $E_0$ and $\omega$ denote the electric field amplitude and the angular frequency of the radiation. $\,\hat{\mathbi{\!e}}$ is the unit polarization vector. The pulse envelope $f(t)$ is defined by the Gaussian-like expression
\begin{equation}
\label{eq:pluse enveloppe}
f(t)=\sin^2\left(\frac{\pi t}{2\tau}\right)
\end{equation}
where $\tau$ is the Full Width at Half Maximum (FWHM), and $2\tau$ the total pulse duration.

Introducing the expansions~(\ref{eq:wf})-(\ref{eq:psi+}) and~(\ref{eq:psiE+}) in the time-dependent Schr\"odinger equation, and projecting onto the electronic and rotational basis functions yields, in the dipole approximation, the following set of coupled differential equations for the nuclear wave packets $\psi^{N_{E}}_{M}(R,t)$ and $\psi^{N^{+}\!,\,\ell}_{M-m}(\varepsilon,R,t)$
\begin{subequations}
\label{eq:tdse}
\begin{eqnarray}
i\hbar\frac{\partial}{\partial t}\psi^{N_{E}}_{M}
& = & \hat{\cal H}^{E}_{N_{E}}\,\psi^{N_{E}}_{M}\nonumber\\
\label{eq:tdseE}
& - & E(t)\times\!\!\!\!\sum_{N^{+}\!,\ell,m\!}\!\!{\cal M}^{N^{+}\!,\,\ell,m}_{N_{E},M}\;\psi^{N^{+}\!,\,\ell}_{M-m}\\
i\hbar\frac{\partial}{\partial t}\psi^{N^{+}\!,\,\ell}_{M-m}
& = & \left(\hat{\cal H}^{+}_{N^+}+\varepsilon\right)\psi^{N^{+}\!,\,\ell}_{M-m}\nonumber\\
\label{eq:tdse+}
& - & E(t)\times\sum_{N_{E}}{\cal M}^{N^{+}\!,\,\ell,m\,^*}_{N_{E},M}\;\psi^{N_{E}}_{M}
\end{eqnarray}
\end{subequations}
where the nuclear Hamiltonian $\hat{\cal H}^{+}_{N^{+}}$ is given by
\begin{equation}
\label{eq:nucham+}
\hat{\cal H}^{+}_{N^{+}} =
-\frac{\hbar^2}{2\mu}\left[\frac{\partial^2}{\partial R^2}-\frac{N^{+}(N^{+}\!+1)}{R^2}\right] + V_{+}(R)
\end{equation}
The matrix elements ${\cal M}^{N^{+}\!,\,\ell,m}_{N_{E},M}$ which couple the nuclear wave packets evolving on the electronic potential curves $V_{E}(R)$ and $V_{+}(R)$ read
\begin{eqnarray}
\label{eq:coupling}
{\cal M}^{N^{+}\!,\,\ell,m}_{N_{E},M}
& = &
\sum_{N}(2N+1)
\left(
\begin{array}{ccc}
N^{+}   & \ell & N \\
M-m & m    & -M
\end{array}
\right)\nonumber\\
& \times &
\left(
\begin{array}{ccc}
N_{E} & 1 & N \\
M     & 0 & -M
\end{array}
\right) \{N_{E}|N|N^{+}\ell\}
\end{eqnarray}
where the total angular momentum (ion + electron)
\begin{equation}
\label{eq:Ntotal}
\vec{\mathbi{N}} = \vec{\mathbi{N}}^{+} + \vec{\boldsymbol{\ell}}
\end{equation}
has been introduced.

The source term $\{N_{E}|N|N^{+}\ell\}$ is evaluated in the molecular frame. The electron wave function originally described in the laboratory frame in Eq.~(\ref{eq:psi+}) is expressed in the molecular frame (Hund's case~(b) representation) using the frame transformation technique implemented by Ugo Fano in his pioneering work on H$_{2}$~\cite{Fano}. Following these lines, we obtain
\begin{eqnarray}
\label{eq:sumLambda}
\{N_{E}|N|N^{+}\ell\}
& = &
\sum_{\Lambda=0}^{1}\;\langle N^{+}\ell | N\Lambda \rangle\;{\mathrm{e}}^{i\pi\mu_\Lambda}\nonumber\\
& &
\qquad\;\times\;d^{\,\ell}_{\Lambda}(\varepsilon,R)\;\langle N_{E}1 | N\Lambda \rangle
\end{eqnarray}
where $\Lambda$ is the projection of the total angular momentum $\vec{\mathbi{N}}$ on the molecular axis.

From the definition given in Eq.~(\ref{eq:Ntotal}), the following relation holds
\begin{equation}
\label{eq:Lambdatotal}
\Lambda = \Lambda^{+} + \lambda
\end{equation}
where $\Lambda^{+}$ and $\lambda$ are the projections of the ion and electron angular momenta $\vec{\mathbi{N}}^{+}$ and $\vec{\boldsymbol{\ell}}$ on the molecular axis. In the present case, the ionic core presents a $^2\Sigma_g^+$ symmetry (\mbox{$\Lambda^{+}=0$}) and hence
\begin{equation}
\label{eq:Lambdalambda}
\Lambda=\lambda
\end{equation}
In Eq.~(\ref{eq:sumLambda}), the phases \mbox{($\pi\mu_\Lambda$)} with \mbox{$\Lambda=0$} \mbox{or 1} represent the phase shifts of the $\sigma$ and $\pi$ electron continuum wave functions relative to the regular radial Coulomb function. The short range quantum defects $\mu_\Lambda$ with $\Lambda=0$ or 1 are indeed associated with the $\Sigma$ and $\Pi$ $p$-Rydberg series of \lit\cite{Fano}.

The matrix element \mbox{$\langle N^{+}\ell | N\Lambda \rangle$} of the unitary frame transformation and the H\"onl-London rotational factor \mbox{$\langle N_{E}1 | N\Lambda \rangle$} of Eq.~(\ref{eq:sumLambda}) are defined by
\begin{eqnarray}
\label{eq:OL}
\langle N'\ell' | N\Lambda \rangle
& = &
(-1)^{N'+\Lambda+1}\,(2-\delta_{\Lambda0})^{\frac{1}{2}}\,(2N'+1)^{\frac{1}{2}}\nonumber\\
& &
\times
\left(
\begin{array}{ccc}
\ell'    & N       & N' \\
-\Lambda & \Lambda & 0
\end{array}
\right)
\end{eqnarray}
In the sum of Eq.~(\ref{eq:sumLambda}) $d^{\,\ell}_{\Lambda}(\varepsilon,R)$ denotes the energy and $R$-dependent ionization dipole moment from the $E$-state
\begin{equation}
\label{eq:dipoletot}
d^{\,\ell}_{\Lambda}(\varepsilon,R) = 
\left(
\begin{array}{ccc}
\ell     & 1       & 0 \\
-\Lambda & \Lambda & 0
\end{array}
\right)
d_{\ell}(\varepsilon,R)
\end{equation}
where
\begin{eqnarray}
\label{eq:dipole}
d_{\ell}(\varepsilon,R)
& = &
\int \phi^{*}_{\ell}(\varepsilon,r|R)\,\Phi_{+}^{*}(\vec{\mathbi{r}}_{\mathbi{\!c\,}}|R)\nonumber\\
&   & \qquad\times\;r\;
\phi_{E}(r,\vec{\mathbi{r}}_{\mathbi{\!c\,}}|R)
\;dr\,d\vec{\mathbi{r}}_{\mathbi{\!c\,}}
\end{eqnarray}
The present study is limited to a restricted range of photoelectron energies \mbox{$(\varepsilon < 200\,{\mathrm{cm}}^{-1})$} and of internuclear distances (in the vicinity of the $E$-state equilibrium distance $R_e$). This justifies the Condon approximation $d_{\ell}(\varepsilon,R) \simeq {\mathrm{cst}}$ used hereafter. The quantum defects $\mu_{\Sigma}$ and $\mu_{\Pi}$ are also taken as independent of $R$, and their numerical values \mbox{$\mu_{\Sigma} \simeq 0.001$} and \mbox{$\mu_{\Pi} \simeq -0.287$} have been extracted at \mbox{$R=R_e$} from the electronic potential energies given in~\cite{schmidt85}.

\subsection{The time propagation}
\label{sec:time propagation}

To calculate the ionization of \lit subjected to a pulsed laser radiation, we propagate the nuclear wave packets $\psi^{N_{E}}_{M}(R,t)$ and $\psi^{N^{+}\!,\,\ell}_{M-m}(\varepsilon,R,t)$ in time during the entire pulse using the split operator method developed by \mbox{Feit {\em et al}}~\cite{feit82}
\begin{equation}
\label{eq:propag}
\left(
\begin{array}{c}
\vdots\\
\psi^{N_{E}}_{M}\\
\vdots\\
\psi^{N^{+}\!,\,\ell}_{M-m}\\
\vdots
\end{array}
\right)_{\!t+\delta t}
= \;\mathrm{e}^{-i\,\hat{\cal H}\,\delta t/\hbar}\,
\left(
\begin{array}{c}
\vdots\\
\psi^{N_{E}}_{M}\\
\vdots\\
\psi^{N^{+}\!,\,\ell}_{M-m}\\
\vdots
\end{array}
\right)_{\!t}
\end{equation}
where the total (molecular + interaction) Hamiltonian $\hat{\cal H}=\hat{T}+\hat{V}+\hat{W}(t)$ is split in three parts corresponding to the kinetic ($\hat{T}$), potential ($\hat{V}$) and interaction ($\hat{W}(t)$) propagators
\begin{eqnarray}
\label{eq:split}
\mathrm{e}^{-i\,\hat{\cal H}\,\delta t/\hbar}
& = &
\mathrm{e}^{-i\,\hat{T}\,\delta t/2\hbar}
\;
\mathrm{e}^{-i\,\hat{V}\,\delta t/2\hbar}
\;
\mathrm{e}^{-i\,\hat{W}(t)\,\delta t/\hbar}\nonumber\\
& &
\times
\mathrm{e}^{-i\,\hat{V}\,\delta t/2\hbar}
\;
\mathrm{e}^{-i\,\hat{T}\,\delta t/2\hbar}
+o(\delta t^3)
\end{eqnarray}
The kinetic and potential propagations are performed in the momentum and coordinate spaces respectively. Fast Fourier Transformation (FFT) allows rapid passage back and forth from one representation to the other at each time step. The propagator associated with the laser interaction term $\hat{W}(t)$ is calculated using a simple diagonalization of its associated interaction matrix~\cite{charron98}.

Since we are only dealing with bound vibrational states in this study, typical grids extend from \mbox{$R=3.5$\ au} to \mbox{$R=14.0$\ au} with $2^{7}$ grid points. The potential energy curves $V_{E}(R)$ and $V_{+}(R)$ are taken from~\cite{schmidt85}. Numerically, the rotating wave approximation (RWA), very accurate for the present case of low laser intensities and vertical resonant transitions, allows a substantial gain of computational time. This approximation, which consists in neglecting the so-called counter-rotating terms, results in a simple one-photon dressing (energy translation by $\hbar\omega$) of the ion electronic potential $V_{+}(R)$ in our time-dependent approach~\cite{charron98}. A time step of the order of \mbox{$\delta t \simeq 4 \,\mathrm{fs} \ll \tau$} is then sufficient for convergence.

\subsection{The photoelectron spectra}
\label{sec:photoelectron spectra}

The analysis of the photoelectron angular distributions is made by projecting the wave packet $\psi_{+}$ defined in Eq.~(\ref{eq:psi+}) at the end of the pulse \mbox{($t=2\tau$)} on the energy-normalized solutions of the field-free ionized molecular states~\cite{continuum}. We need to define the set of outgoing plane waves elastically scattered in the direction \mbox{$\vec{\mathbi{k}} \equiv (k,\,\hat{\mathbi{\!k}}\,)$} for a prescribed asymptotic electron kinetic energy \mbox{$\varepsilon=\hbar^{2}k^{2}/2m$}. These are represented by the usual expansions on angular momentum states~\cite{Messiah}
\begin{equation}
\label{eq:continuum}
|\varepsilon,\,\hat{\mathbi{\!k}}\,\rangle = \sum_{\ell,m}\,i^{\ell}\,\mathrm{e}^{-i\xi_{\ell}}\,
Y_{\ell m}^{*}(\,\hat{\mathbi{\!k}}\,)\,\phi_{\ell}(\varepsilon,r|R)\,Y_{\ell m}(\,\hat{\mathbi{\!r}}\,)
\end{equation}
where $\xi_{\ell}$ denotes the Coulomb phase shift of each partial wave. We therefore evaluate, for a given value of $M$, the angular distribution of the ejected photoelectron at some fixed energy $\varepsilon$ as
\begin{equation}
\label{eq:P_M(E,theta)}
P_{M}(\varepsilon,\,\hat{\mathbi{\!k}}\,)=
\int\!dR\,d\,\hat{\mathbi{\!R}}\;
\left|\,\langle\,\varepsilon,\,\hat{\mathbi{\!k}}\;
|\,\psi_{+}(\vec{\mathbi{r}},\vec{\mathbi{R}},2\tau)\,\rangle_{\vec{\mathbi{r}}}\right|^{2}
\end{equation}
This multiple integral can be written in a more convenient form with the help of the expansions~(\ref{eq:psi+}) and~(\ref{eq:psi+lm}). After the integration over the electronic coordinate \mbox{$\vec{\mathbi{r}} \equiv (r,\,\hat{\mathbi{\!r}}\,)$} and over the angle \mbox{$\,\hat{\mathbi{\!R}}\, \equiv (\theta_{R},\phi_{R})$} we get
\begin{eqnarray}
\label{eq:P(E,theta)simple}
P_{M}(\varepsilon,\,\hat{\mathbi{\!k}}\,)
& = &
\!\!\sum_{N^{+}}\,\sum_{m}\,\sum_{\ell,\ell'}
i^{(\ell'-\ell)}\,\mathrm{e}^{i(\xi_{\ell}-\xi_{\ell'})}
Y_{\ell' m}^{*}(\,\hat{\mathbi{\!k}}\,)
Y_{\ell m}(\,\hat{\mathbi{\!k}}\,)\nonumber\\
& \times &
\!\!\!\!\int\!\!dR\,\,
\psi^{N^{+}\!,\,\ell'^{\,*}}_{M-m}(\varepsilon,R,2\tau)\;
\psi^{N^{+}\!,\,\ell}_{M-m}(\varepsilon,R,2\tau)
\end{eqnarray}
One can notice here the appearance of an incoherent sum over the quantum numbers $N^{+}$ and $m$, while the sum over the electron angular momentum $\ell$ is coherent and gives rise to various (\mbox{$\ell\times\ell'$}) cross-terms. The total photoelectron angular distribution at some fixed energy $\varepsilon$ is then obtained by averaging over the initial distribution of $M$
\begin{equation}
\label{eq:sumM}
P(\varepsilon,\,\hat{\mathbi{\!k}}\,) \propto \sum_{M}P_{M}(\varepsilon,\,\hat{\mathbi{\!k}}\,)
\end{equation}

In the present study, a $p$-electron is ejected and the value of the electron angular momentum is therefore fixed to \mbox{$\ell=1$}. In this case, Eqs.~(\ref{eq:P(E,theta)simple}) and~(\ref{eq:sumM}) simplify to
\begin{eqnarray}
\label{eq:P(E,theta)simple2}
P(\varepsilon,\,\hat{\mathbi{\!k}}\,)
& \propto &
\sum_{M}\,\sum_{N^{+}}\sum_{m}\;\left|Y_{1m}(\,\hat{\mathbi{\!k}}\,)\right|^2\nonumber\\
&         &
\qquad\times\;\int\,\left|\psi^{N^{+},1}_{M-m}(\varepsilon,R,2\tau)\right|^{2}dR
\end{eqnarray}
The total photoelectron spectrum is then obtained by a summation over the ejection angle \mbox{$\,\hat{\mathbi{\!k}}\, \equiv (\theta_{k},\phi_{k})$}
\begin{equation}
\label{eq:P(E)}
P(\varepsilon) = \int\!d\,\hat{\mathbi{\!k}}\; P(\varepsilon,\,\hat{\mathbi{\!k}}\,)
\end{equation}
thus giving
\begin{equation}
\label{eq:P(E)simple}
P(\varepsilon) \propto
\sum_{M}\,\sum_{N^{+}}\sum_{m}\int\,\left|\psi^{N^{+},1}_{M-m}(\varepsilon,R,2\tau)\right|^{2}dR
\end{equation}
Numerically, the continuous variable $\varepsilon$ is discretized in 150 energy values, with \mbox{$10 \leqslant \varepsilon \leqslant 190\,{\mathrm{cm}}^{-1}$}. The probability that the electron exits in the $\hat{\mathbi{\!k}}$ direction with an energy $\varepsilon$ is calculated from Equation~(\ref{eq:P(E,theta)simple2}) on a grid of 325~points in \mbox{$\,\hat{\mathbi{\!k}}\, \equiv (\theta_{k},\phi_{k})$}.

Even if  Eq.~(\ref{eq:P(E)simple}) only reveals a series of incoherent sums, one should not forget that each exit channel $N^+$ may be reached from different initial levels $N_E$. As a consequence, an initial quantum superposition of rotational levels can induce an interference effect in the photoelectron spectra which arises from the phases of the different components in this initial wave packet.

\section{Results and discussion}
\label{sec:Results and discussion}

\subsection{Analysis of the photoelectron spectra and of the interference effect}

\begin{figure}[!ht]
\includegraphics[width=8cm,clip]{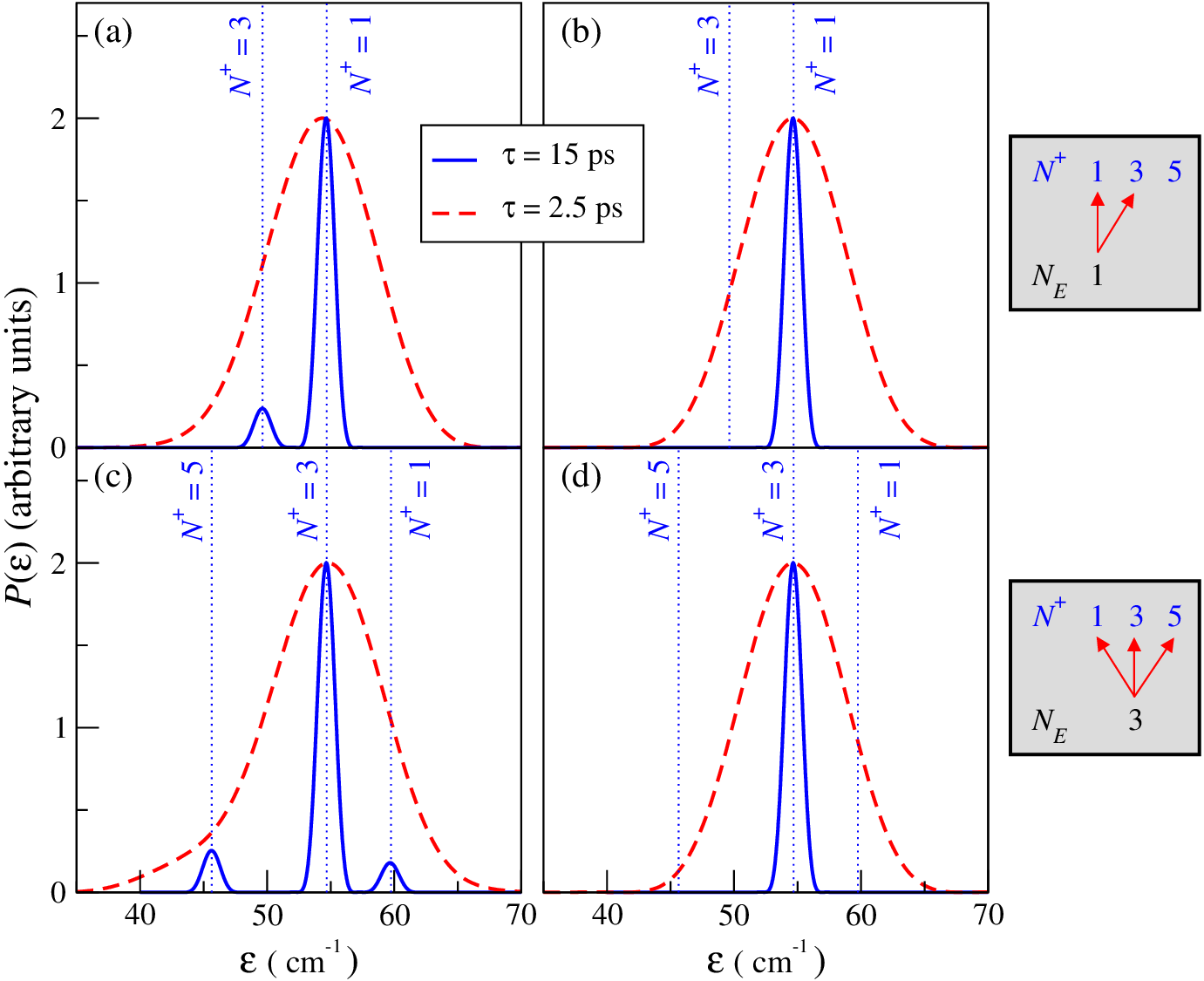}
\caption{
\label{fig:f2}
(Color online) Photoelectron spectra calculated using a {\em single\/} initial state \mbox{$(v_{E},N_{E})$} as a function of energy $\varepsilon$. Upper panels (a) and (b): the initial state is \mbox{$v_{E}=0$}, \mbox{$N_{E}=1$}. Lower panels (c) and (d): the initial state is \mbox{$v_{E}=0$}, \mbox{$N_{E}=3$}. The photoelectron spectra calculated using the quantum defects of \lit are shown in the left panels (a) and (c), while the spectra in the right panels (b) and (d) correspond to \mbox{$\mu_\Sigma=\mu_\Pi$} (see text for details). The solid blue lines correspond to the pulse duration (FWHM) \mbox{$\tau = 15\ \mathrm{ps}$} while \mbox{$\tau = 2.5\ \mathrm{ps}$} for the red dashed lines. The laser wavelength is \mbox{705\ nm}. The photoelectron energies expected from a simple energy conservation rule \mbox{$\varepsilon=E(v_{E},N_{E})+\hbar\omega-E(v_{+},N^{+})$} with  the usual selection rule \mbox{$v_{+}=v_{E}$} are shown as thin vertical dotted blue lines for various ion rotational quantum numbers $N^{+}$.
}
\end{figure}

Typical photoelectron spectra are shown Figure~\ref{fig:f2} when a single initial rovibrational level \mbox{($v_E,N_E$)} is populated. The upper and lower graphs of Figure~\ref{fig:f2} correspond to the same initial vibrational state \mbox{$v_E=0$}, but to different initial rotational excitations: \mbox{$N_E=1$} in the upper panels (a) and (b), \mbox{$N_E=3$} in the lower panels (c) and (d). It is assumed here that these levels are prepared using the two-step process depicted Figure~\ref{fig:pot}, with \mbox{($v_X=0, N_X=1$)} and \mbox{($v_A=0, N_A=2$)}. The incoherent sum over $M$ in Eqs.~(\ref{eq:P(E,theta)simple2}) and~(\ref{eq:P(E)simple}) therefore extends from \mbox{$M=-1$} to \mbox{$M=1$} only. The photoelectron spectra shown in the left panels (a) and (c) have been calculated using the accurate short range quantum defects \mbox{$\mu_\Sigma=0.001$} and  \mbox{$\mu_\Pi=-0.287$} of \lit while the right panels (b) and (d) correspond to an hypothetically isotropic \lit molecule with  \mbox{$\mu_\Sigma=\mu_\Pi$}.

The solid blue lines in Figure~\ref{fig:f2} represent the photoelectron spectra obtained with the laser frequency \mbox{$\omega=14184$\ cm$^{-1}$} (wavelength \mbox{705\ nm}) and with the pulse duration \mbox{$\tau=15$\ ps}. One can notice that for the initial condition \mbox{$N_E=1$} (panel (a)) two peaks are seen in the photoelectron signal. Their energies correspond to the ion exit channels \mbox{$v_+=0$} with \mbox{$N^+=1$} or 3.

The \mbox{$v_+=v_E$} vibrational selection rule is observed in this case because of the pronounced Rydberg character of the $E$-electronic state at short internuclear distances. The two potential curves \mbox{$V_E(R)$} and \mbox{$V_+(R)$} are indeed parallel for \mbox{$R \leqslant 7$\ au} (see Figure~\ref{fig:pot}). Around \mbox{$R \simeq 9$\ au}, an avoided crossing gives rise to the double-well structure of the $E$-state, deeply modifying the Rydberg nature of this excited state. However, this avoided crossing has a negligible influence on the lowest vibrational levels (\mbox{$v_E \leqslant 4$}).

The branching ratio between the \mbox{$N^+ = 1$} and \mbox{$N^+ = 3$} exit channels clearly favors \mbox{$N^+ = N_E = 1$}. This result is not unexpected since in the case of an isotropic symmetry (corresponding here to \mbox{$\mu_\Sigma=\mu_\Pi$}) the other ionization channel (\mbox{$N^+ = 3$}) has no intensity, as one can see in the panel (b) of Figure~\ref{fig:f2}. The molecular anisotropy, expressed by the phase difference \mbox{$(\mu_\Sigma-\mu_\Pi)\pi \simeq 0.3\pi$}, allows for the exchange of angular momentum between the ionized electron and the nuclear rotation. This effect, which can be seen as a mutual electron-nuclei entanglement, explains the appearance of the additional branch \mbox{$N^+ = 3$} in panel (a).

For the initial level \mbox{$N_E=3$} (lower part of Figure~\ref{fig:f2}), this effect is also seen with the appearance of the two satellite peaks \mbox{$N^+ = N_E \pm 2$} around \mbox{$\varepsilon=45$ cm$^{-1}$} and around \mbox{$\varepsilon=60$ cm$^{-1}$}. These two peaks are not seen in panel (d) when \mbox{$\mu_\Sigma=\mu_\Pi$}. Note also from panels (a) and (c) that the central peak, which corresponds to an ionization without exchange of angular momentum (\mbox{$N^+ = N_E$}), is located at the same photoelectron energy \mbox{$\varepsilon=55$ cm$^{-1}$} for both initial rotational levels \mbox{$N_E=1$} and~3. This happens because the rotational constants of the $E$-state of \lit and of the ground electronic state of \litp are almost identical.

The dashed red lines shown Figure~\ref{fig:f2} finally represent the photospectra calculated with the same parameters except for a much shorter pulse duration $\tau=2.5$~ps. In this case the peaks associated with different values of the ion rotational quantum number $N^+$ overlap due to the large spectral bandwidth of the pulse. A single broad photoelectron peak is, therefore, obtained around \mbox{$\varepsilon=55$ cm$^{-1}$} whatever the initial rotational level $N_E$.

In the isotropic case shown on the right hand side of Figure~\ref{fig:f2} the photoelectron peaks calculated with this shorter pulse duration have a symmetric shape since a single ionization channel (\mbox{$N^+ = N_E$}) is observed. On the other hand, a slightly asymmetric shape is obtained with the real \lit molecule (see the red dashed line in panel (c) for instance), due to the asymmetric distribution of the two satellite peaks \mbox{$N^+ = N_E \pm 2$} on both sides of the central peak corresponding to \mbox{$N^+ = N_E$}.

\begin{figure}[!ht]
\includegraphics[width=8cm,clip]{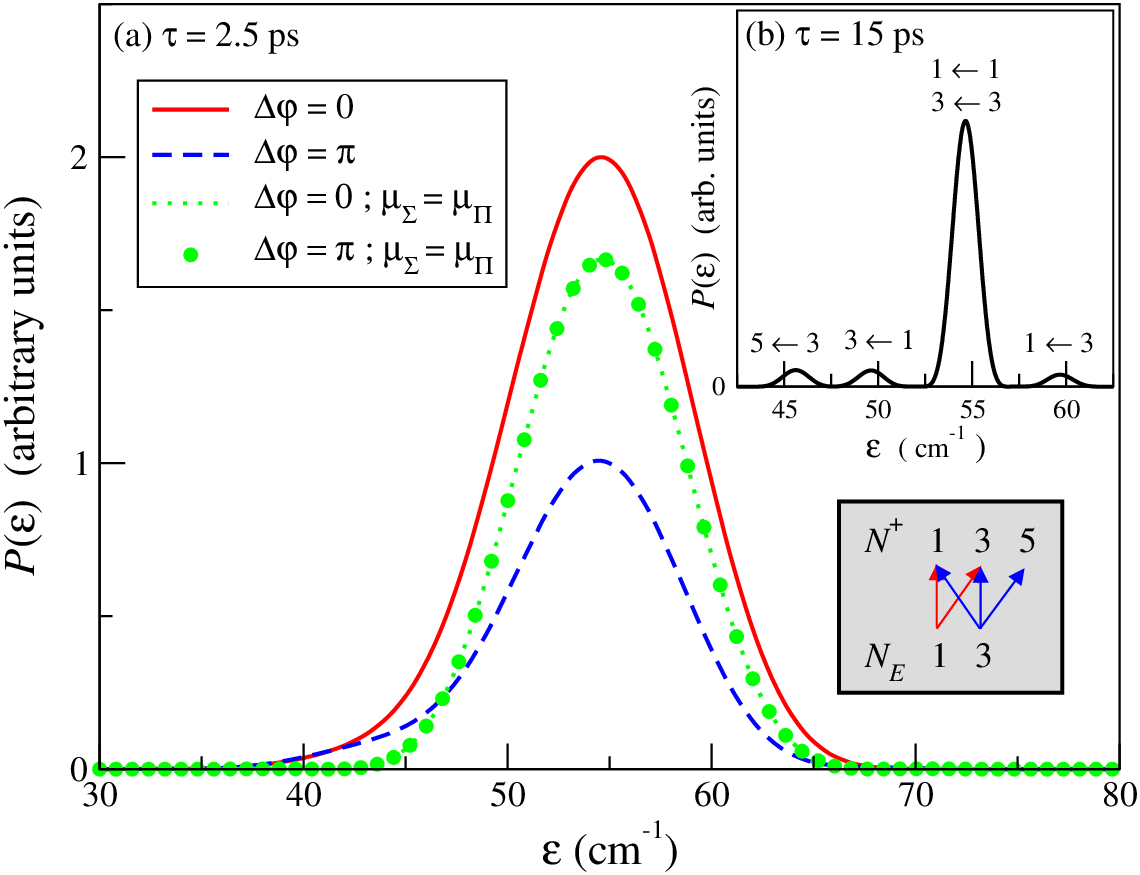}
\caption{
\label{fig:f3}
(Color online) Photoelectron spectra calculated using a {\em single\/} initial vibrational state (\mbox{$v_{E}=0$}) with a coherent superposition of {\em two\/} rotational levels (\mbox{$N_{E}=1$} and \mbox{$N_{E}=3$}) as a function of energy $\varepsilon$. The pulse duration is \mbox{$\tau = 2.5\ \mathrm{ps}$} in the main graph (a) and \mbox{$\tau = 15\ \mathrm{ps}$} in the small upper-right inset (b). The laser wavelength is \mbox{705\ nm}. The red solid line and blue dashed line correspond to an {\em in phase\/} \mbox{$(\Delta\varphi=0)$} and an {\em out of phase\/} \mbox{$(\Delta\varphi=\pi)$} coherent superposition respectively. The isotropic case \mbox{$\mu_\Sigma=\mu_\Pi$} is represented by the green dotted line for \mbox{$\Delta\varphi=0$} and the green solid circles for \mbox{$\Delta\varphi=\pi$}. The labels \mbox{$N^{+} \!\leftarrow\! N_{E}$} in the inset assign the various energy peaks with respect to the initial ($N_{E}$) and final ($N^{+}$) rotational levels. See text for details.
}
\end{figure}

In the experiment, it is expected that two $N_E$ rotational levels can be populated in a coherent distribution whose initial phase difference is controlled using the phase mask of a pulse shaper. This type of experiment has been implemented recently for higher rotational levels in the Group of Stephen Leone~\cite{Leone2} for instance. The photoelectron spectra calculated using in-phase and out-of-phase initial distributions of \mbox{$N_{E}=1$} and 3 are shown in Figure~\ref{fig:f3} as red solid and blue dashed lines for the pulse duration \mbox{$\tau = 2.5\ \mathrm{ps}$}. One can recognize in these two spectra the slightly asymmetric distributions discussed previously.

A clear and significant interference effect is also seen here. The total ionization probability is amplified by a factor of 2.05 when comparing the in-phase and out-of-phase initial conditions.

Because of the orthogonality of the Wigner rotation matrices ${{\cal D}^{N^{\,*}}_{M,\Lambda}}(\,\hat{\mathbi{\!R}}\,)$ the photoelectron probability can be written as an incoherent sum over the rotational quantum number $N^+$, as one can see in Eq.~(\ref{eq:P(E)simple}). The interference effect seen here is therefore not due to an interference between the different exit channels labeled by $N^{+}$, but is due to an interference taking place in the same ionization channel between the two pathways coming from the two possible initial levels $N_E$.

Let us consider for instance the ionization channel associated with $N^+=3$. This final state can be reached through the main branch coming from $N_E=3$ or through the small satellite arising from $N_E=1$. One could think at first sight that the cross term associated with this interference mechanism is probably negligible considering the small branching ratio (\mbox{$\simeq 0.1$}) between the \mbox{$N^+ = N_E$} and \mbox{$N^+ = N_E \pm 2$} pathways (see Figure~\ref{fig:f2}c). However, even in this apparently unfavorable case, it can be easily estimated that this cross term can induce a significant interference effect which could in principle modify the ionization probability by a factor of three when comparing destructive and constructive interferences.

In our case all exit channels are not subjected to this interference. For example \mbox{$N^+ = 5$} can be reached from \mbox{$N_E = 3$} only. In addition the photoelectron signal is a complex incoherent average over the different possible values of $M$ and $m$ (see Eq~(\ref{eq:P(E)simple})). The two different pathways \mbox{$N_E \rightarrow N^+$} and \mbox{$N_{E}' \rightarrow N^+$} are also affected by the different values of the coupling matrix elements ${\cal M}^{N^{+}\!,\,\ell,m}_{N_{E},M}$ given in Eq.~(\ref{eq:coupling}). The contrast separating a constructive from a destructive interference is therefore not maximum, and after this complex averaging Figure~\ref{fig:f3} shows that the interference effect changes the ionization probability of \lit by a factor of about two.

\begin{figure}[!ht]
\includegraphics[width=8cm,clip]{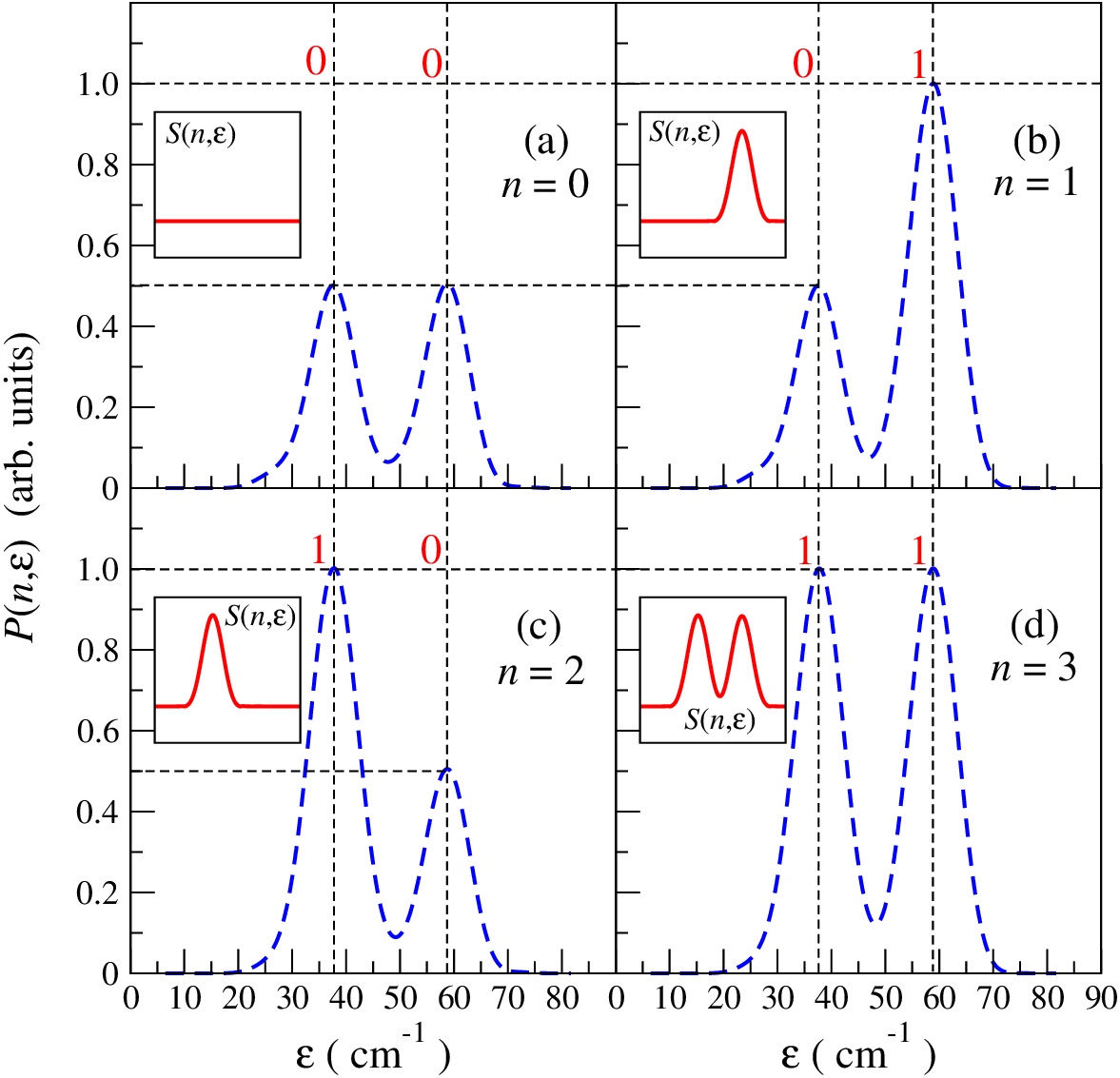}
\caption{
\label{fig:f4}
(Color online) Photoelectron spectra calculated using {\em two\/} initial vibrational states (\mbox{$v_{E}=0$} and 1) with a coherent superposition of {\em two\/} rotational levels (\mbox{$N_{E}=1$} and \mbox{$N_{E}=3$}) as a function of energy $\varepsilon$. The pulse duration is \mbox{$\tau = 2.5\ \mathrm{ps}$} and the laser wavelength is 699.8~nm. The dashed blue lines show the photoelectron spectra $P(n,\varepsilon)$ calculated when the integer values $n=0$ \mbox{(panel a)}, \mbox{1 (panel b)}, \mbox{2 (panel c)} and \mbox{3 (panel d)} are stored in the phase difference between the \mbox{$N_{E} = N_{A} \pm 1$} rotational components of the initial wave packet. The solid red lines in the small insets show the signal difference \mbox{$S(n,\varepsilon) = P(n,\varepsilon) - P(0,\varepsilon)$} for \mbox{$n=0\ldots3$}.
}
\end{figure}

Our analysis is also confirmed by the photoelectron spectrum obtained with \mbox{$\mu_\Sigma=\mu_\Pi$} (green dotted line and green solid circles in Figure~\ref{fig:f3}). Whatever the phase shift \mbox{$\Delta\varphi$} between the \mbox{$N_{E}=1$} and \mbox{$N_{E}=3$} initial rotational levels the same photoelectron spectrum is obtained, and no interference effect is seen. This is due to the disappearance of the \mbox{$N_E \rightarrow N^+ = N_E \pm 2$} pathways, as discussed previously.

The interference effect can also be suppressed by increasing the pulse duration $\tau$. In this last case the photoelectron peaks associated with different rotational quantum number $N^+$ do not overlap anymore, and two identical $N^+$ originating from different $N_E$ never appear at the same energy (see the small inset in Figure~\ref{fig:f3} for instance). As can be seen in Eq.~(\ref{eq:P(E)simple}), the wave packets associated with different photoelectron energies never interfere and as a consequence the interference effect seen with shorter pulses is not observed with long pulses. We have indeed verified that the spectrum shown in the small inset of Figure~\ref{fig:f3} (\mbox{$\tau=15$ ps}) does not depend on \mbox{$\Delta\varphi$}.

We will now show that the interference effect shown in Figure~\ref{fig:f3} with relatively short pulses can be used to reveal efficiently a more complex phase information initially stored in the rotational wave packet of \lit in a way similar to the implementation performed in the group of Philip Bucksbaum with Rydberg wave packets~\cite{Bucksbaum1,Bucksbaum2}.

\subsection{Applications to quantum information}

In \lit, we use the quantum superposition of two rotational levels in different vibrational states to store a binary information. The case of two vibrational levels is shown in Figure~\ref{fig:f4}. The combinations 00, 01, 10 and 11, associated with the decimal values $n=0$, 1, 2 and 3, can then be stored. The measurement of this information is performed by photoionization, thanks to the anharmonicity of the electronic potential curves of the molecule.

Since the vibrational frequency of the $E$-state and of the ion are almost identical, in an harmonic approximation and with \mbox{$v_+=v_E$} the energy of the ionized electron
\begin{subequations}
\begin{eqnarray}
\label{eq:energy}
\varepsilon &    =   & \!\!E(v_{E},N_{E})+\hbar\omega-E(v_{+},N^{+})\\
            & \stackrel{\textrm{\scriptsize harm}}{\simeq} & \!\!\mathrm{cst} + B_{\rm rot}\left[N_E(N_E\!+\!1)-N_+(N_+\!+\!1)\right]\;
\end{eqnarray}
\end{subequations}
does not depend on the value of the initial vibrational level. However, the anharmonicity of \lit is large enough (\mbox{$\simeq 20-50$ cm$^{-1}$}) to exceed by far the rotational spacing (\mbox{$B_{\rm rot} \simeq 0.5$ cm$^{-1}$}). Different initial vibrational levels can therefore be clearly distinguished even with the relatively short pulses (\mbox{$1\,\mathrm{ps} \leqslant \tau \leqslant 5\,\mathrm{ps}$}) which allow for the observation of the interference effect discussed previously.

The dashed blue lines in Figure~\ref{fig:f4} represent the photoelectron spectra calculated with the phase differences (0,0), (0,$\pi$), ($\pi$,0) and ($\pi$,$\pi$) in the vibrational states \mbox{($v_{E}=1$,$v_{E}=0$)} between the rotational levels \mbox{$N_{E}=1$} and \mbox{$N_{E}=3$}. These phase differences correspond to the integers $n=0$, 1, 2 and 3. We denote by $P(n,\varepsilon)$ the associated photoelectron spectra. The interference effect (amplification of the ionization probability by a factor of two) is clearly seen for both vibrational levels.

If we define the signal difference $S(n,\varepsilon)$ by the following relation
\begin{equation}
\label{eq:S(n,E)}
S(n,\varepsilon) = P(n,\varepsilon) - P(0,\varepsilon)\,,
\end{equation}
a direct visualization of the binary representation of $n$ is obtained in the graphs showing $S(n,\varepsilon)$ as a function of the energy $\varepsilon$. These signal differences are represented as red solid lines in the four small insets of Figure~\ref{fig:f4}. This photoelectron signal difference $S(n,\varepsilon)$ characterizes the cross terms associated with the \mbox{$N^+=N_E$} and \mbox{$N^+=N_{E}'\pm2$} paths. It can also be seen as a direct measure of the electron-nuclei entanglement induced by the molecular anisotropy.

\begin{figure}[!ht]
\includegraphics[width=8cm,clip]{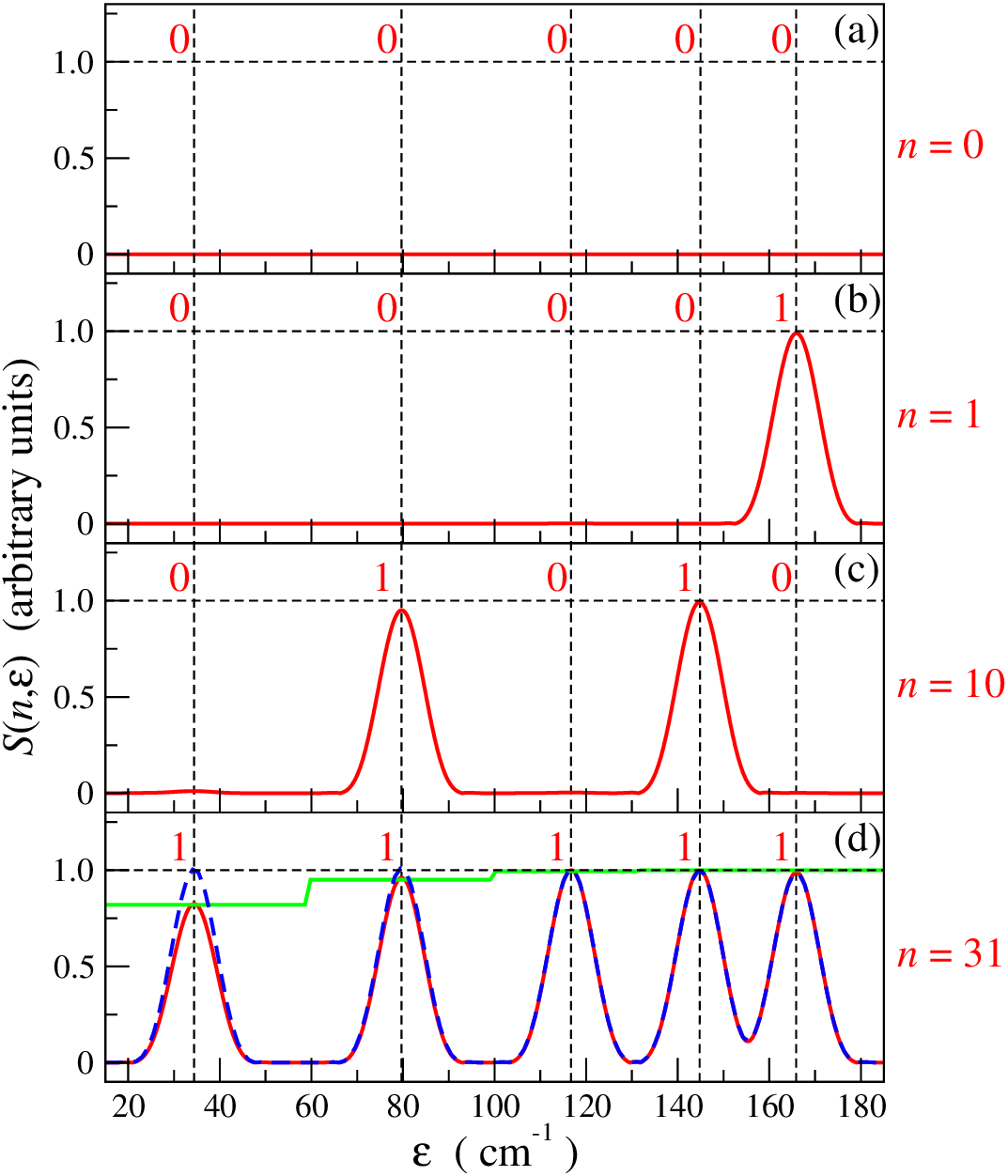}
\caption{
\label{fig:f5}
(Color online) Photoelectron spectra calculated using {\em five\/} initial vibrational states (\mbox{$v_{E}=0\ldots4$}) with a coherent superposition of {\em two\/} rotational levels (\mbox{$N_{E}=1$} and \mbox{$N_{E}=3$}) as a function of energy $\varepsilon$. The pulse duration is \mbox{$\tau = 2.5\ \mathrm{ps}$} and the laser wavelength is \mbox{699.6\ nm}. The solid red lines show the signal difference \mbox{$S(n,\varepsilon) = P(n,\varepsilon) - P(0,\varepsilon)$} when the integer values \mbox{$n=0$}~(panel a) , 1~(panel b), 10~(panel c) and 31~(panel d) are stored in the phase difference between the \mbox{$N_{E} = N_{A} \pm 1$} rotational components of the initial wave packet. The step function shown as a green solid line in panel (d) represents the vibrational Franck-Condon factors \mbox{$\mathscr{F}(\varepsilon)=\langle v_{+} | v_{E} \rangle$} associated with the different photoelectron peaks. The ratio \mbox{$S(n,\varepsilon)/\mathscr{F}(\varepsilon)$} is shown in panel (d) as a blue dashed-line.
}
\end{figure}

In order to check the scalability of this method, we have calculated the signal difference $S(n,\varepsilon)$ for $n=0$, 1, 10 and 31 when the binary information is stored in the first five vibrational levels of the $E$-state. These results are shown as red solid lines in the four panels of Figure~\ref{fig:f5}. One can notice the effectiveness of the method for such a small number of vibrational levels. 

A few tendencies are worth being noted in this figure. Since the total (electron+ion) energy is fixed, the lowest energy peaks can be assigned to the highest ion vibrational levels $v_+$. The fact that the energy separation between two peaks increases with $v_+$ is simply explained by the increasing anharmonicity of the potential curve with larger internuclear distances.

Finally, one can also notice that the energy peak on the left is not as intense as the other ones. This peak corresponds to an ion left in \mbox{$v_+=4$} and therefore to the molecule being initially in \mbox{$v_E=4$}. This initial level is already located at an energy close to the avoided crossing of the $E$-state (see Figure~\ref{fig:pot}), and the \mbox{$v_+=v_E$} selection rule is no longer fully verified. This initial level is effectively also ionized in \mbox{$v_+=v_E-1=3$}, yielding the ejection of an electron carrying much more energy. This effect explains the smallest probability observed in Figure~\ref{fig:f5} for the \mbox{$v_E = 4 \rightarrow v_+ = 4$} pathway.

Higher vibrational levels also show this tendency of being ionized into vibrational levels with \mbox{$v_+ < v_E$}. This type of behavior could be seen as a limitation of the proposed mechanism for storing and reading information in and from the rotational degree of freedom of a diatomic molecule. However, this effect can easily be corrected for by renormalizing the photoelectron signal with the vibrational Franck-Condon factors
\begin{equation}
\label{FCF}
\mathscr{F}(\varepsilon) = \int \chi_{v_{+}}^*\!(R)\;\chi_{v_{E}}\!(R)\;dR
\end{equation}
between the eigenfunctions associated with the $v_E$ and \mbox{$v_+=v_E$} vibrational quantum numbers. The step-like function \mbox{$\mathscr{F}(\varepsilon)$} is shown as a green solid line in panel~(d) of Figure~\ref{fig:f5}. The decreasing probability seen in the left peak associated with \mbox{$v_+ = v_E = 4$} is compensated by calculating the simple ratio \mbox{$S(n,\varepsilon)/\mathscr{F}(\varepsilon)$}. This ratio is shown in the same graph as a blue dashed line, and all constructive interference peaks now reach the same height. Since the vibrational wave functions \mbox{$\chi_{v_{E}}\!(R)$} and \mbox{$\chi_{v_{+}}\!(R)$} can be easily calculated from the potential curves given in~\cite{schmidt85}, this correcting procedure is quite straitforward.

\section{Conclusion}
\label{sec:Conclusion}
We have proposed here a theoretical model for the study of the short-pulse photoionization of \lit, following a scheme relatively close to recent experimental implementations. This model takes into account the molecular rotational degree of freedom and allows for the calculation of the photoelectron spectrum resolved in time, energy and in angle.

We have used this time-dependent model to predict a new and efficient mechanism for measuring binary classical information initially stored in the phases of a rotational wave packet. A picosecond laser excitation induces an interference between these different rotational components. This typically molecular interference effect measures the electron-nuclei entanglement which takes place thanks to the anisotropy of the diatomic molecule, and transfers the rotational phase information in the photoelectron spectrum.

These results indicate that a high degree of control can be achieved in this type of molecular systems using simple pulse shaping techniques. A systematic exploration of the control achievable in the photoelectron angular distributions will be presented in another paper.

\begin{acknowledgments}
We thank Herv\'e Le Rouzo and Georges Raseev (Orsay) for stimulating and helpful discussions. The IDRIS-CNRS supercomputer center provided computational time under project number 08/051848. This work was partially supported by the LRC of the CEA, under contract number DSM-05–33. Laboratoire de Photophysique Mol\'eculaire and Laboratoire Aim\'e Cotton are associated with Universit\'e Paris-Sud 11.
\end{acknowledgments}


\begin{thebibliography}{99}

\bibitem{QI}
A. Steane, Rep. Prog. Phys. {\bf 61}, 117 (1998);
C. Bennet and D. DiVincenzo, Nature {\bf 404}, 247 (2000);
P. Shor, {\em Proc. 35th Annual Symp. on Found. of Comput. Science\/}, IEEE Comput. Soc. Press, Santa Fe, NM (1994);
M. A. Nielsen and I. L. Chuang, {\em Quantum Computation and Quantum Information}, Cambridge University Press, Cambridge (2000).

\bibitem{Lloyd99}
S. Lloyd, \pra {\bf 61}, 010301(R) (2000).

\bibitem{Divincenzo00}
P. Divincenzo, Fortschr. Phys. {\bf 48}, 771 (2000).

\bibitem{Jones01}
J. Jones, Prog. Nuc. Magn. Res. Spect., {\bf 38}, 325 (2001).

\bibitem{TI}
J. I. Cirac and P. Zoller, \prl {\bf 74}, 4091 (1995);
C. Monroe, D. M. Meekhof, B. E. King, W. M. Itano, and D. J. Wineland, \prl {\bf 75}, 4714 (1995);
J. I. Cirac and P. Zoller, Nature {\bf 404}, 579 (2000).

\bibitem{QED}
Q. A. Turchette, C. J. Hood, W. Lange, H. Mabuchi, and H. J. Kimble, \prl {\bf 75}, 4710 (1995);
A. Rauschenbeutel, G. Nogues, S. Osnaghi, P. Bertet, M. Brune, J. M. Raimond, and S. Haroche, \prl {\bf 83}, 5166 (1999).

\bibitem{JJ}
A. Shnirman, G. Sch\"on, and Z. Hermon, \prl {\bf 79}, 2371 (1997);
Y. Makhlin, G. Sch\"on, and A. Shnirman, \rmp {\bf 73}, 357 (2001).

\bibitem{NA}
D. Jaksch, H. J. Briegel, J. I. Cirac, C. W. Gardiner, and P. Zoller, \prl {\bf 82}, 1975 (1999);
G. K. Brennen, C. M. Caves, P. S. Jessen, and I. H. Deutsch, \prl {\bf 82}, 1060 (1999);
A. Hemmerich, \pra {\bf 60}, 943 (1999);
E. Charron, E. Tiesinga, F. Mies, and C. Williams, \prl {\bf 88}, 077901 (2002);
O. Mandel, M. Greiner, A. Widera, T. Rom, T. W. H\"ansch, and I. Bloch, Nature {\bf 425}, 937 (2003);
J. V. Porto, S. Rolston, B. Laburthe Tolra, C. J. Williams, and W. D. Phillips, Phil. Trans. R. Soc. Lond. A {\bf 361}, 1417 (2003);
E. Charron, M. A. Cirone, A. Negretti, J. Schmiedmayer, and T. Calarco, \pra, {\bf 74}, 012308 (2006).

\bibitem{MWI}
E. A. Shapiro, M. Spanner, and M. Y. Ivanov, \prl {\bf 91}, 237901 (2003);
E. A. Shapiro, I. Khavkine, M. Spanner, and M. Y. Ivanov, \pra {\bf 67}, 013406 (2003);
J. Degert, C. Meier, B. Chatel, and B. Girard, \pra {\bf 67}, 041402(R) (2003);
K. Ohmori, Y. Sato, E. E. Nikitin, and S. A. Rice, \prl {\bf 91}, 243003 (2003);
K. Ohmori, H. Katsuki, H. Chiba, M. Honda, Y. Hagihara, K. Fujiwara, Y. Sato, and K. Ueda, \prl {\bf 96}, 093002 (2006).

\bibitem{Bucksbaum1}
J. Ahn, T. C. Weinacht, and P. H. Bucksbaum, Science {\bf 287}, 463 (2000);
D. A. Meyer; P. G. Kwiat and R. J. Hughes; P. H. Bucksbaum, J. Ahn, and T. C. Weinacht, Science {\bf 289}, 1431a (2000).

\bibitem{Bucksbaum2}
J. Ahn, D. N. Hutchinson, C. Rangan, and P. H. Bucksbaum, \prl {\bf 86}, 1179 (2001);
C. Rangan, and P. H. Bucksbaum, \pra {\bf 64}, 033417 (2001);
J. Ahn, C. Rangan, D. N. Hutchinson, and P. H. Bucksbaum, \pra {\bf 66}, 022312 (2002).

\bibitem{Leone1}
J. M. Papanikolas, R. M. Williams, P. D. Kleiber, J. L. Hart, C. Brink, S. D. Price, and S. R. Leone, \jcp {\bf 103}, 7269 (1995);
R. Uberna, M. Khalil, R. M. Williams, J. M. Papanikolas, and S. R. Leone, \jcp {\bf 108}, 9259 (1998);
R. Uberna, Z. Amitay, C. X. W. Qian, and S. R. Leone, \jcp {\bf 114}, 10311 (2001);
J. B. Ballard, X. Dai, A. N. Arrowsmith, L. H\"uwel, H. U. Stauffer, and S. R. Leone, Chem. Phys. Lett. {\bf 402}, 27 (2005);
X. Dai, E. A. Torres, E. W. Lerch, D. J. Wilson, and S. R. Leone, Chem. Phys. Lett. {\bf 402}, 126 (2005).

\bibitem{Leone2}
J. B. Ballard, H. U. Stauffer, E. Mirowski, and S. R. Leone, \pra {\bf 66}, 043402 (2002);
J. B. Ballard, A. N. Arrowsmith, L. H\"uwel, X. Dai, and S. R. Leone, \pra {\bf 68}, 043409 (2003).

\bibitem{Leone3}
J. M. Papanikolas, R. M. Williams, and S. R. Leone, \jcp {\bf 107}, 4172 (1997);
H. U. Stauffer, J. B. Ballard, Z. Amitay, and S. R. Leone, \jcp {\bf 116}, 946 (2002);
J. B. Ballard, H. U. Stauffer, Z. Amitay, and S. R. Leone, \jcp {\bf 116}, 1350 (2002);
Z. Amitay, J. B. Ballard, H. U. Stauffer, and  S. R. Leone, Chem. Phys. {\bf 267}, 141 (2001).

\bibitem{Leone4}
Z. Amitay, R. Kosloff, and S. R. Leone, Chem. Phys. Lett. {\bf 359}, 8 (2002);
J. Vala, Z. Amitay, B. Zhang, S. R. Leone, and R. Kosloff, \pra {\bf 66}, 062316 (2002).

\bibitem{Note1}
The fact that the two different rotational levels evolve with a different time dependence ${\mathrm{e}}^{-i\,E_{N_{E}}t/\hbar}$ due to their different energy $E_{N_{E}} \simeq B_{\rm rot}N_{E}(N_{E}+1)$ adds a small complexity which can be dealt with by precompensating this future phase evolution in the initial phase difference \mbox{$\Delta\varphi$} between the two rotational components of the wave packet (see~\cite{Bucksbaum1} for details).

\bibitem{FTPD}
S. C. Althorpe and T. Seideman, \jcp {\bf 110}, 147 (1999);
Y. Arasaki, K. Takatsuka, K. Wang, and V. McKoy, Chem. Phys. Lett. {\bf 302}, 363 (1999);
Y. Arasaki, K. Takatsuka, K. Wang, and V. McKoy, \jcp {\bf 112}, 8871 (2000);
L. Pesce, Z. Amitay, R. Uberna, S. R. Leone, M. Ratner, and R. Kosloff, \jcp {\bf 114}, 1259 (2001).

\bibitem{Zare}
R. N. Zare, {\em Angular momentum\/}, John Wiley \& Sons, New-York (1988).

\bibitem{Fano}
U. Fano, \pra {\bf 2}, 353 (1970);
U. Fano and D. Dill, \pra {\bf 6}, 185 (1972);
D. Dill, \pra {\bf 6}, 160 (1972);
C. Jungen and O. Atabek, \jcp {\bf 66}, 5584 (1977);
C. Jungen and D. Dill, \jcp {\bf 73}, 3338 (1980);
M. Raoult and C. Jungen, \jcp {\bf 74}, 3388 (1981).

\bibitem{schmidt85}
I. Schmidt-Mink, W. M\"uller, and W. Meyer, Chem. Phys. {\bf 92}, 263 (1985).

\bibitem{feit82}
M. J. Feit, J. A. Fleck, and A. Steiger, J. Comput. Phys. {\bf 47}, 412 (1982).

\bibitem{charron98}
E. Charron and A. Suzor-Weiner, \jcp {\bf 108}, 3922 (1998).

\bibitem{continuum}
E. Charron, A. Giusti-Suzor, and F. H. Mies, \pra {\bf 49}, R641 (1994).
X. Chen, A. Sanpera, and K. Burnett, \pra {\bf 51}, 4824 (1995).

\bibitem{Messiah}
A. Messiah, {\em Quantum Mechanics\/}, Vol. I, North Holland, Amsterdam (1962).

\end{thebibliography}
\end{document}